\documentclass[fleqn]{article}
\usepackage{espcrc2}
\usepackage{epsfig}

\title{Status of and performance estimates for QCDOC \thanks{Presented
by PAB at Lattice 2002, Boston.}}

\author{
P.A.~Boyle\address[ED]{School of Physics, University 
  of Edinburgh, Edinburgh EH9 3JZ, Scotland}\address[Columbia]{Department 
  of Physics, Columbia University, New York, NY, 10027},
D.~Chen\address[IBM]{IBM T.J.~Watson Research Center, Yorktown
  Heights, NY, 10598}, 
N.H.~Christ\addressmark[Columbia], 
C.~Cristian\addressmark[Columbia], 
Z.~Dong\addressmark[Columbia],
A.~Gara\addressmark[IBM],
B.~Jo\'o\addressmark[Ed]\addressmark[Columbia],
C.~Jung\addressmark[Columbia],
C.~Kim\addressmark[Columbia], 
L.~Levkova\addressmark[Columbia], 
X.~Liao\addressmark[Columbia],
G.~Liu\addressmark[Columbia],
R.D.~Mawhinney\addressmark[Columbia],
S.~Ohta\address[RBRC]{Institute for Particle and Nuclear Studies,
  KEK, Tsukuba, Ibaraki, 305-0801, Japan}\address[RBNL]{RIKEN-BNL 
  Research Center, Brookhaven National Laboratory, Upton, NY, 11973},
K.~Petrov\addressmark[Columbia],
T.~Wettig\addressmark[RBNL]\address{Department of Physics, Yale
  University, New Haven, CT, 06520-8120},
A.~Yamaguchi\addressmark[Columbia]}
     
\begin{document}

\begin{abstract}
QCDOC is a supercomputer designed for high scalability at 
a low cost per node.
We discuss the status of the project and provide performance
estimates for large machines obtained from cycle accurate simulation
of the QCDOC ASIC.

\end{abstract}

\maketitle

\section*{Introduction}
\vspace{-.3cm}
QCDOC\cite{Lat00,Lat01} was designed
for a cost-effective balance between memory bandwidth,
floating point performance and communication performance
when running massively parallelised double precision QCD codes.
As shall be seen, our simulations suggest
QCDOC should scale to exceptionally large machine sizes
on a fixed scientifically interesting lattice volume - or equivalently
QCDOC will operate efficiently on very small local volumes.
This gives the focussed compute power necessary to simulate much lighter
quark masses than are accessible today, at a cost of 1 US-dollar per sustained
Megaflop.

\vspace{-.3cm}
\section*{Hardware Overview}
\vspace{-.3cm}
The QCDOC design is based on an application specific integrated 
circuit, or ASIC.
We use IBM System-On-a-Chip technology to integrate all 
the node logic on a single silicon chip. 
The CPU core is a PowerPC 440\cite{440Core} whose FPU\cite{Dockser01} 
can execute one operate instruction per clock cycle. Peak performance 
of two flops per clock is obtained by using fused multiply-add
instructions.
The on-chip features include 4MByte of embedded DRAM memory
and its controller, a memory controller for off-chip DDR memory, 
an Ethernet interface, and a custom serial communications unit (SCU)
linked by a number of on-chip busses.

The prefetching edram controller (PEC) has been custom 
designed by our colleagues at IBM Research, and provides ample bandwidth
for the FPU. It automatically
prefetches two read streams, and is multi-ported
allowing remote communication to overlap local computation 
without contention.

The SCU hardware implements a 6 dimensional hyper-torus, with 
bi-directional low latency links. 
The aggregate 1.5GByte/s inter-node bandwidth easily supports QCD with 
1 Gflop of local floating point all the way down to $2^4$ local volumes. 
The DMA engines in the SCU transfer complex patterns between nodes while
the local floating point calculation proceeds uninterrupted.

The high dimensionality of the network 
both increases the communication bandwidth of the machine, and 
gives very symmetric local volumes when distributing the problem
over very many nodes in four, or even five, dimensions. This
yields the most favorable surface to volume ratio. 

\begin{table}[ht]
\caption{
\label{tabAsm}
Performance of double precision assembler kernels produced
using the speaker's code generator.
Very high fractions of the 1Gflop raw peak, and even higher fractions
of the theoretical peak, are obtained. 
Some compiled code figures are included to show that
reasonable performance can be obtained without undue pain.
}
\begin{tabular}{c|c}
Operation & Mflops/node \\
\hline
SU3-SU3      &  800\\
SU3-2spinor        &  780\\
DAXPY              &  190\\
ZAXPY              &  450\\
DAXPY-Norm         &  350\\
CloverTerm/asm     &  790\\
CloverTerm/gcc     &  150\\
CloverTerm/xlc     &  300
\end{tabular}
\end{table}

\begin{table}[ht]
\caption{
\label{tabDslash}
Sample performance of red-black fermion operators in cycle accurate
simulation. The Wilson and Clover kernels were generated
using the speaker's C++ code scheduler, while the Staggered operator
was hand coded by Calin Cristian. The codes are
very efficient even when all sites are on multiple boundaries.
}
\begin{tabular}{c|c|c}
Operation  & Local Vol. &  Mflops/node\\
\hline
Wilson    $D_{eo}$ & $2^4$ & 470\\
Wilson    $D_{eo}$ & $4^4$ & 535\\
Clover    $D_{eo}$ & $2^4$ & 560\\
Clover    $D_{eo}$ & $4^4$ & 590\\
Staggered $D_{eo}$ & $2^4$ & 370\\
Staggered $D_{eo}$ & $2^2.4^2$ & 430
\end{tabular}
\end{table}

\begin{figure}[ht]
\caption{
Estimated performance per node for 
assembly coded Wilson dslash operators as a function of local volume. 
Roughly two orders of magnitude better scalability is observed for QCDOC.
}
\label{figScaling} 
\epsfig{file=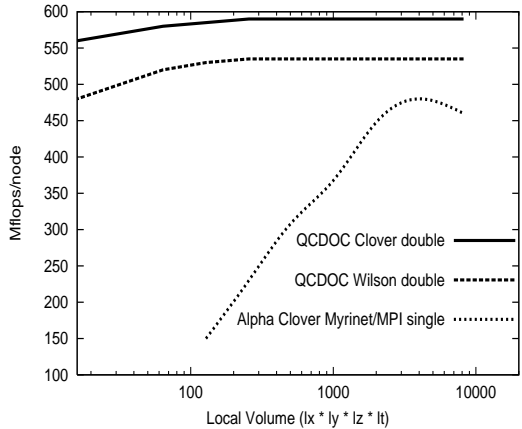,height=6cm,width=7cm}
\end{figure}

\begin{table}[ht]
\caption{
\label{tabScaling}
Estimates for Wilson CG performance on a $32^3\times 64$ Lattice.
Both Clover and Domain Wall simulations would be even more scalable.
}
\begin{tabular}{c|c|c|c}
Nodes & $M^\dag M$  & Gsum & Sust. Tflops\\
\hline
4096 & 2620$\mu$s & 10  $\mu$s & 2.15\\
8192 & 1310$\mu$s & 11.5$\mu$s & 4.2\\
16384& 680 $\mu$s & 13  $\mu$s & 8.1\\
32768& 340 $\mu$s & 15  $\mu$s & 15.6\\
\end{tabular}
\end{table}

\vspace{-.3cm}
\section*{Hardware Status}
\vspace{-.3cm}
The QCDOC ASIC design is functionally complete, and the
preliminary design release is undergoing timing driven physical layout 
at IBM Raleigh.  Verification is on-going with
release to manufacture expected by mid fall and first silicon
at the end of this year. 
Large prototype machines are expected in spring 2003.

\vspace{-.3cm}
\section*{Software Environment}
\vspace{-.3cm}
Two widespread and standard compilers are being used, namely the IBM xlc 
compiler and the GNU gcc C and C++ compilers. Debug facilities are 
provided by a port of the full featured RISCWatch remote debug tool.

The standard runtime support libraries are a popular GNU Public License
libc released by Cygnus\footnote{Popularised in the Cygwin software
package}. Runtime libraries are available for performing
communication over the physics network via the SciDac QMP interface
which is in turn implemented on top of a native SCU interface.

Assembler optimised QCD kernels will be freely available.

\vspace{-.3cm}
\section*{ Node kernels}
\vspace{-.3cm}
The use of the PowerPC means that standard operating systems
could in principle be run on the nodes. However, to scale to very small
local volumes it is essential that, in addition to very low latency hardware,
we avoid unnecessary software latency.

We will use a lean node kernel to avoid scheduler overhead and use
the PowerPC virtual memory (VM) hardware to protect, but not
translate, memory pages. This gives both the benefits of 
robust and graceful error recovery, and the benefits of
zero-copy SCU transfers without VM 
gymnastics.


\vspace{-.3cm}
\section*{ Host operating system}
\vspace{-.3cm}
The host for QCDOC will be an SMP Unix server with multiple 
Gigabit links to the Boot/Diagnostic/IO Ethernet network.
A multi-threaded qdaemon will boot and manage the machine partitions,
and service socket-based connections to these partitions from a number of
client programs. Sufficient functionality has already been implemented
to boot and download code to PowerPC boards in parallel.

\vspace{-.3cm}
\section*{ Performance in simulation}
\vspace{-.3cm}
As part of the design verification programme a number of 
benchmarks and stress tests have been written, to both check
functional correctness and that design goals are met.
We shall present performance figures for some of these tests,
where we have taken a ``nominal'' 500MHz CPU, and made reasonable
assumptions about the interconnect wire length.

Table~\ref{tabAsm} shows a sample of common vector operations
performed in Lattice QCD codes. Most of these figures were
obtained using a C++ assembler code generator written by the 
speaker to automate loop unrolling, prefetching and detailed
scheduling. 

Table~\ref{tabDslash} shows performance measurements of 
optimised implementations of various preconditioned Fermion operators
in double precision. In these operators unpaired adds and multiplies 
set an upper limit (e.g. 780Mflops for the Wilson kernel) somewhat
below the ``raw'' peak.
The communication overhead is entirely taken into account, 
and the performance is excellent despite having to communicate
each site over four wires in the $2^4$ case. 

\vspace{-.3cm}
\section*{Scalability}
\vspace{-.3cm}

Performance characteristics at small local volumes 
are critical to scaling. Figure~\ref{figScaling} shows the
estimated performance of QCDOC on double precision Wilson dslash code 
(omits global summation time) distributed in four dimensions. 
At the smallest volumes the entire
operation takes roughly 20 $\mu$s, with eight communications
in this time. For comparison we overlay 
a Myrinet Alpha 21264 cluster running single precision assembler code 
distributed in three dimensions.

Given the fast linear algebra and the scalable matrix multiply
the remaining hurdle to be overcome by an iterative solver is 
global summation. 
The SCU has ``pass-thru'' hardware assist for global sums and broadcasts.
The global sum has been benchmarked in simulation, and we
use the mixture of matrix multiplies, linalg and gsums in the
CG algorithm to predict the performance on large machines
on Wilson HMC in Table~\ref{tabScaling}.

\vspace{-.3cm}
\section*{Conclusions}
\vspace{-.3cm}

QCDOC is progressing well with large machines due in 2003. 
Simulations indicate these machines will run QCD at high
efficiency on the largest machines, while having a low
cost per sustained Megaflop and low power consumption. 

\vspace{-.3cm}
\section*{Acknowledgements}
\vspace{-.3cm}
This research was supported in part by the U.S. Department of Energy,
the Institute of Physical and Chemical Research (RIKEN) of Japan, and
the U.K. Particle Physics and Astronomy Research Council.

\end{document}